\documentclass[11pt,letterpaper]{article}
\usepackage{systeme} 
\usepackage[utf8]{inputenc}
\usepackage{multirow} 
\usepackage{booktabs} 
\usepackage[english]{babel}
\usepackage{amsmath}
\usepackage{amsfonts}
\usepackage{amssymb}
\usepackage{graphicx}
\usepackage[small,bf]{caption} 
\usepackage[left=3cm,right=3cm,top=2cm,bottom=3cm]{geometry}
\usepackage{framed} 
\usepackage{color} 
\usepackage{wrapfig}\definecolor{shadecolor}{RGB}{220,220,220} 
\usepackage{float} 
\usepackage{array} 
\usepackage{caption} 

\author{Jorge Pinochet}
\title{\textbf{Black hole evaporation: From frontier physics to the classroom}}
\begin{document}

\author{Jorge Pinochet$^{*}$\\ \\
 \small{$^{*}$\textit{Departamento de Física. Centro de Desarrollo de Investigación CEDI-UMCE,}}\\
 \small{\textit{Universidad Metropolitana de Ciencias de la Educación,}}\\
 \small{\textit{Av. José Pedro Alessandri 774, Ñuñoa, Santiago, Chile.}}\\
 \small{e-mail: jorge.pinochet@umce.cl}\\}

\date{}
\maketitle

\begin{center}\rule{0.9\textwidth}{0.1mm} \end{center}
\begin{abstract}
\noindent Half a century ago, Stephen Hawking discovered that black holes have a temperature, emit thermal radiation, and evaporate. Although Hawking temperature has been widely discussed in this journal, the evaporation process has received less attention. This work presents an accessible analysis of black hole evaporation, addressing the main questions raised by this phenomenon through intuitive ideas and heuristic calculations. It is shown that a problem situated at the frontier of physics can be reformulated in terms accessible for exploration in the classroom. In this sense, the article constitutes an educational resource that enables the integration of concepts from thermodynamics, relativity, and quantum mechanics within a meaningful learning context.\\ \\


\begin{center}\rule{0.9\textwidth}{0.1mm} \end{center}
\end{abstract}

\maketitle

\section{Introduction}

The sensational discovery made by Stephen Hawking in 1974 ---that ``black holes ain’t so black''--- marked a milestone in the development of twentieth-century physics [1]. Hawking's central result can be summarized in a single statement: black holes behave as hot bodies with an absolute temperature ---the \textit{Hawking temperature}--- that depends inversely on their mass [2–3]. The most immediate and striking consequence of this discovery is that black holes evaporate, eventually disappearing in a burst of gamma rays [3].\\

Although several articles have addressed black holes from an educational perspective, including their tidal effects, basic concepts, common misconceptions, and the teaching of gravitational waves associated with black hole mergers [4–7], other contributions have focused specifically on Hawking radiation and Hawking’s contributions to black-hole physics [8–15]. In particular, LoPresto [16] presented simple thermodynamic estimates of black-hole evaporation. Nevertheless, the evaporation process has received comparatively little attention in the educational literature. Questions such as whether a black hole can evaporate completely, which particles are emitted during evaporation, how much power it radiates, and how long the process takes remain suitable for an introductory treatment. The aim of this work is to address these questions through intuitive arguments and heuristic calculations, making this frontier topic accessible to students and teachers without requiring advanced mathematics. Beyond obtaining numerical estimates, this approach aims to help students explore the physical relationships underlying black-hole temperature, radiation power, and evaporation time, while recognizing the limitations of a heuristic treatment.\\

Sections~2 and~3 provide an overview of the topic, introducing the concepts of black holes, Hawking radiation, and Hawking temperature. Section~4 analyzes the power radiated by a black hole, presents a heuristic derivation of the evaporation time, and discusses the possibility of empirically verifying Hawking's discovery. Section~5 studies particle production during evaporation, and Section~6 examines the final stages of the process. The article concludes with a brief summary and commentary.

\section{Schwarzschild black hole}

The concept of a black hole emerged within the framework of general relativity, which is the theory of gravity proposed by Einstein to extend and refine Newton's law of universal gravitation [17–18]. A black hole forms when a high concentration of mass, or its equivalent in energy, occurs within a closed spherical region of space known as the \textit{event horizon}. In the case of a non-rotating and electrically neutral black hole, commonly referred to as a \textit{Schwarzschild black hole}\footnote{According to an important result known as the \textit{no-hair theorem}, there are only three parameters that characterize a black hole: mass, electric charge, and angular momentum. Therefore, the simplest black hole is the Schwarzschild black hole, which is characterized by mass alone.} (Fig.~1), the radius of the event horizon is calculated using an equation found in 1916 by the German astronomer Karl Schwarzschild [19]:

\begin{equation}
R_S = \frac{2GM_{BH}}{c^2} \simeq 3\,\text{km} \left(\frac{M_{BH}}{M_{\odot}}\right).
\end{equation}

This is the \textit{Schwarzschild radius}, where $G = 6.67 \times 10^{-11}\,\text{N}\cdot\text{m}^2\cdot\text{kg}^{-2}$ is the universal gravitational constant, $c = 3 \times 10^{8}\,\text{m}\cdot\text{s}^{-1}$ is the speed of light in vacuum, and $M_{BH}$ is the mass of the black hole, that is, the mass confined within the event horizon. Following a common practice in astrophysics, $R_S$ has been expressed in units of solar mass, $M_{\odot} = 1.99 \times 10^{30}\,\text{kg}$, and the equation is written in kilometers ($km$). This notation allows quick numerical estimates and will also be used in the following sections\footnote{To express the Schwarzschild radius in terms of the solar mass, one only needs to multiply the right-hand side of the equation by $1.99 \times 10^{30}\,\text{kg}/M_{\odot} = 1$ and substitute the numerical values of $G$ and $c$. In the following sections, an analogous procedure will be used to express other physical quantities in terms of the solar mass.}. To gain an idea of the colossal mass-energy concentrations involved, note that if in Eq.~(1) we take $M_{BH} = M_{\odot}$, then $R_S \simeq 3\,\text{km}$, which corresponds to one hundred-thousandth of the solar radius. Nevertheless, it should be clarified that the Sun will never become a black hole, since it does not have sufficient mass for that to occur.\\

According to general relativity, inside the event horizon lies a central \textit{singularity}, where all the mass of the black hole is compressed into a mathematical point of zero size and infinite density. Due to the colossal gravity generated by the singularity, no form of matter or energy can cross the event horizon from the inside to the outside, because doing so would require a speed greater than that of light in vacuum, which is forbidden by the laws of relativistic physics [19–20]. The event horizon can only be crossed from the exterior to the interior. For our purposes, the most important consequence of this is that the event horizon appears completely black to an external observer. As a result, for this same observer the absolute temperature of the black hole is zero, since any body at a temperature above $0\,\text{K}$ must emit radiation. This astonishing result led the physicist John Archibald Wheeler to coin the term ``black hole'' in 1967 [20].

\begin{figure}[H]
  \centering
    \includegraphics[width=0.35\textwidth]{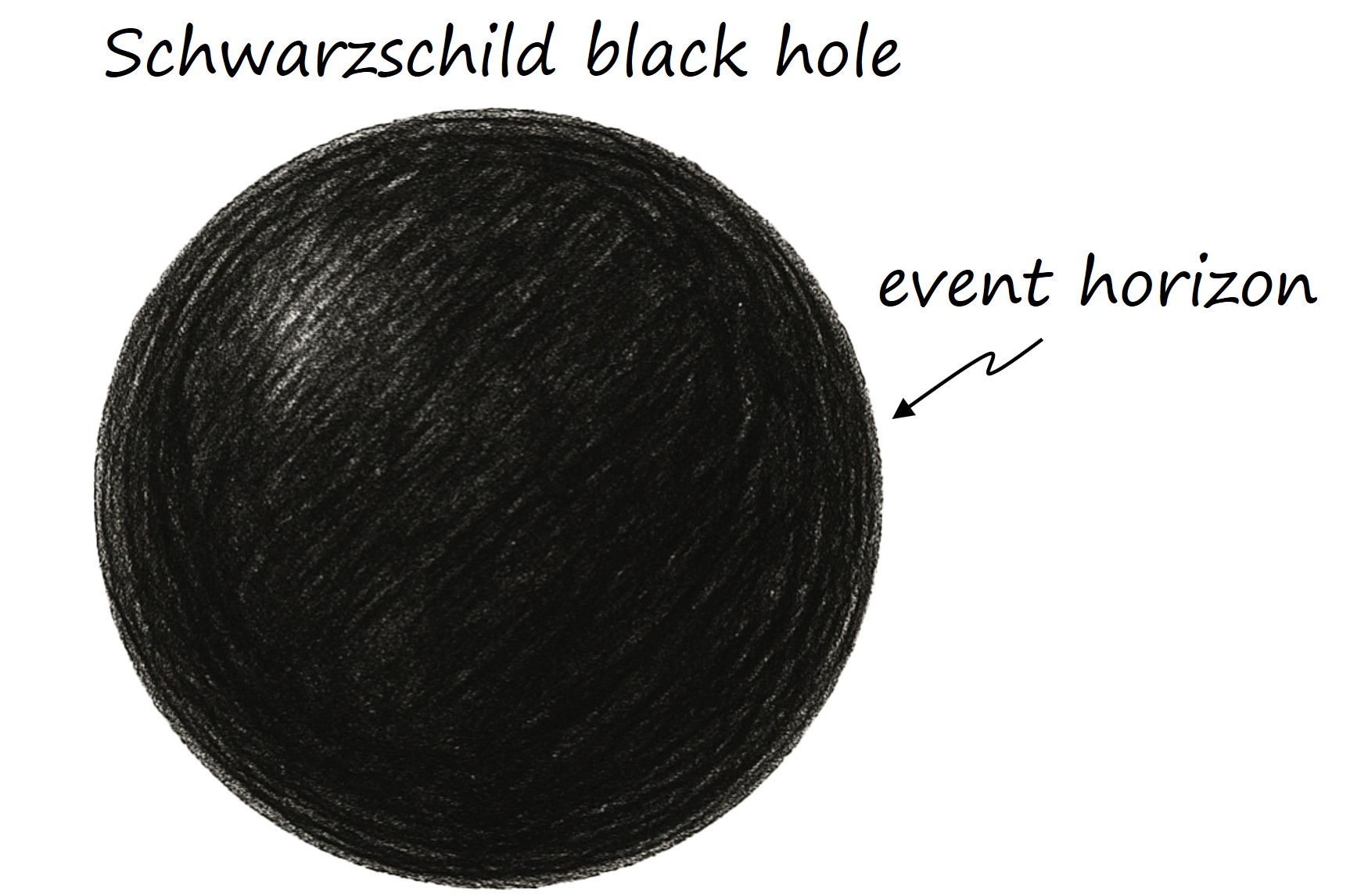}
  \caption{Schematic representation of a Schwarzschild black hole and its event horizon.}
\end{figure}

\section{Hawking radiation and Hawking temperature}

In the previous section, we presented the classical paradigm of a black hole, based solely on general relativity. However, by combining Einstein's theory with quantum mechanics, Hawking made the remarkable discovery that black holes have temperature and emit thermal radiation. This emission is known as \textit{Hawking radiation}, and, for an observer located far from the event horizon, it exhibits a blackbody spectrum\footnote{A rigorous derivation of Hawking radiation requires quantum field theory in curved spacetime. The heuristic arguments presented here are intended to explore the physical consequences of this phenomenon rather than to derive its underlying mechanism.}. These ideas allow one to obtain an estimate of the Hawking temperature through a simple heuristic argument [13]. According to Wien's displacement law, the absolute temperature $T$ of a blackbody and the wavelength $\lambda$ at which the maximum emission occurs are related by the equation [21]:

\begin{equation}
\lambda T = \frac{2\pi \hbar c}{5k} \simeq 2.90 \times 10^{-3}\,\text{m}\cdot\text{K},
\end{equation}

where $\hbar = h/2\pi = 1.05 \times 10^{-34}\,\text{J}\cdot\text{s}$ is the reduced Planck constant and $k = 1.38 \times 10^{-23}\,\text{J}\cdot\text{K}^{-1}$ is the Boltzmann constant. Since the Schwarzschild radius is the characteristic length scale of a Schwarzschild black hole, assuming that this object behaves as a blackbody, we may take $\lambda \approx R_S = 2GM_{BH}/c^2$. Substituting this result into Wien's law, we obtain

\begin{equation}
T \approx \frac{\hbar c^3}{(5/\pi) k G M_{BH}}.
\end{equation}

This equation reveals that $T$ is inversely proportional to $M_{BH}$. Apart from dimensionless factors, this result is identical to the exact expression for the Hawking temperature [22]:

\begin{equation}
T_H = \frac{\hbar c^3}{8\pi k G M_{BH}} = 6.17 \times 10^{-8}\,\text{K} \left(\frac{M_{\odot}}{M_{BH}}\right).
\end{equation}

Here, $T_H$ has been expressed in units of solar mass, and the equation is given in kelvin (K). As this equation shows, and as illustrated in Fig.~2, smaller and lighter black holes have higher temperatures. For example, if we consider a black hole of solar mass, $M_{BH} = M_{\odot}$, we find that $T_H = 6.17 \times 10^{-8}\,\text{K}$, a temperature very close to absolute zero and undetectable through astronomical observations. Since the lightest black holes detected so far have masses of the order of the solar mass (stellar black holes), these results suggest that the quantum effects predicted by Hawking are unobservable with current technology.\\

\begin{figure}[H]
  \centering
    \includegraphics[width=0.4\textwidth]{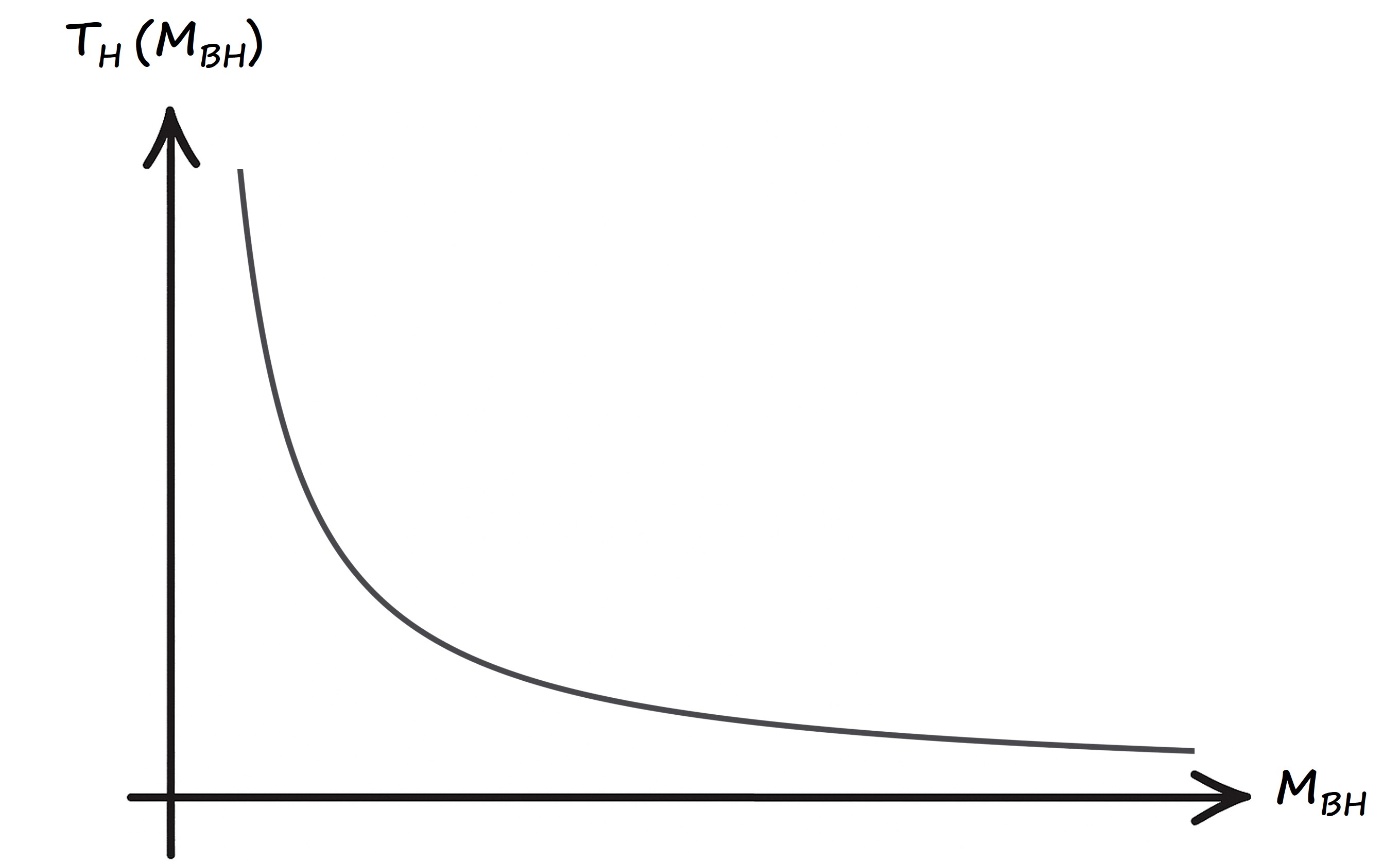}
  \caption{Plot of the Hawking temperature versus black hole mass.}
\end{figure}

However, if we consider a hypothetical \textit{micro black hole} with a mass comparable to that of a mountain, $M_{BH} = 10^{12}\,\text{kg}$, Eq.~(4) shows that $T_H = 1.2 \times 10^{11}\,\text{K}$, an extremely high temperature which, according to Wien's law, corresponds to a peak emission in the gamma-ray region, $\lambda \sim 10^{-14}\,\text{m}$. The problem is that there is no observational evidence for the existence of micro black holes.\\

\begin{figure}[H]
  \centering
    \includegraphics[width=0.65\textwidth]{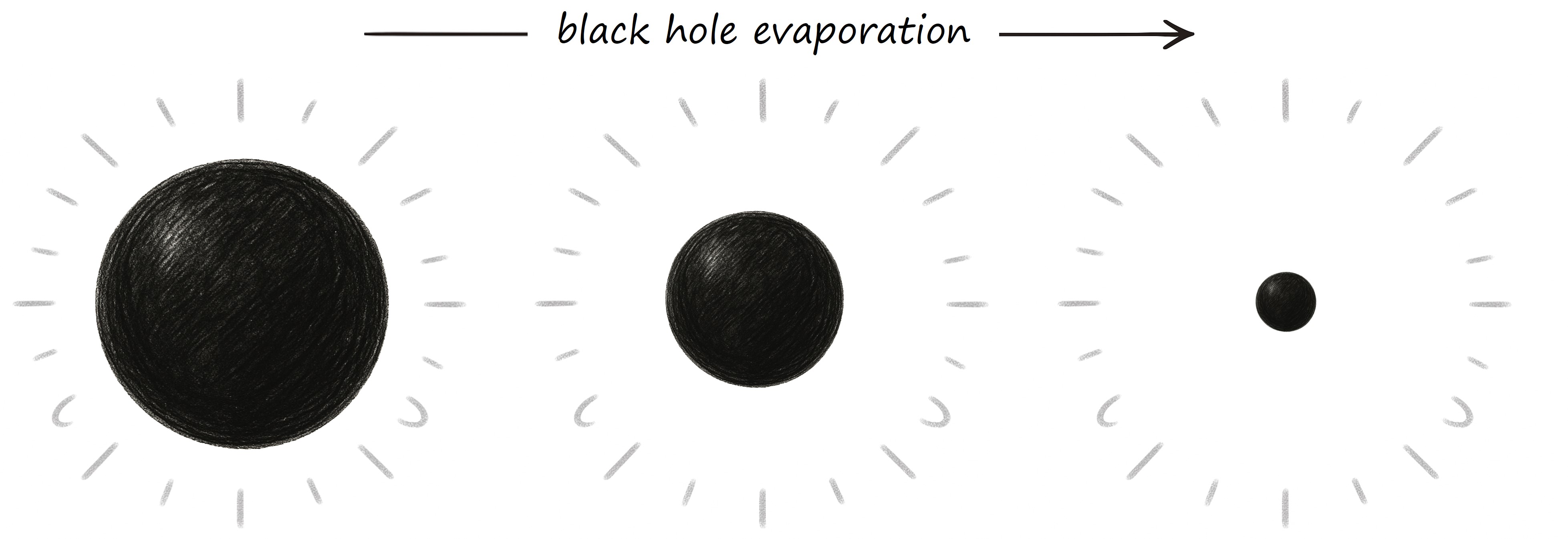}
  \caption{As a black hole evaporates, its mass and radius decrease gradually.}
\end{figure}

Equation (4) shows that as \(M_{BH}\) decreases, \(T_H\) increases, and the radiation becomes more intense. Thus, \(M_{BH}\) decreases more and more rapidly in a process known as evaporation. The time interval that elapses until \(M_{BH}=0\) is known as the \textit{evaporation time} [14]. A black hole loses mass as it emits Hawking radiation (Fig.~3). As will be discussed in detail in the next section, the particle species contributing to this radiation depend on the Hawking temperature: if \(T_H\) is sufficiently high, or \(M_{BH}\) is sufficiently small, particles other than photons can also be emitted, such as neutrinos, electrons, and protons.

\section{Radiated power and evaporation time}

It is possible to obtain a simple estimate of the evaporation time by assuming that the black hole behaves as a blackbody whose power (energy radiated per unit time) follows the Stefan--Boltzmann law [23]:

\begin{equation}
P = \sigma A T_H^4,
\end{equation}

where $\sigma = 5.67 \times 10^{-8}\,\text{W}\cdot\text{m}^{-2}\cdot\text{K}^{-4}$ is the Stefan--Boltzmann constant, and $A$ is the area of the event horizon, which, according to Eq.~(1), is given by

\begin{equation}
A = 4\pi R_S^2 = \frac{16\pi G^2 M_{BH}^2}{c^4}.
\end{equation}

Substituting into Eq.~(5) the values of $T_H$ and $A$ given by Eqs.~(4) and (6), we obtain

\begin{equation}
P = \frac{c^8 \hbar^4 \sigma}{256\pi^3 G^2 k^4 M_{BH}^2}.
\end{equation}

Considering that $\sigma = \pi^2 k^4 / (60 \hbar^3 c^2)$ [24], this equation can be simplified:

\begin{equation}
P = \frac{c^6 \hbar}{15360\pi G^2 M_{BH}^2} = 9 \times 10^{-29}\,\text{W} \left(\frac{M_{\odot}}{M_{BH}}\right)^2.
\end{equation}

As illustrated in the graph in Fig.~4, the smaller $M_{BH}$ is, the greater the power and the faster the black hole radiates energy outward. For example, for a black hole of solar mass, $M_{BH} = M_{\odot}$, the radiated power is very low, $P = 9 \times 10^{-29}\,\text{W}$, but for a hypothetical micro black hole with a mass of $10^{12}\,\text{kg}$, the power is much higher, $P = 3.6 \times 10^{8}\,\text{W}$.\\

\begin{figure}[H]
  \centering
    \includegraphics[width=0.4\textwidth]{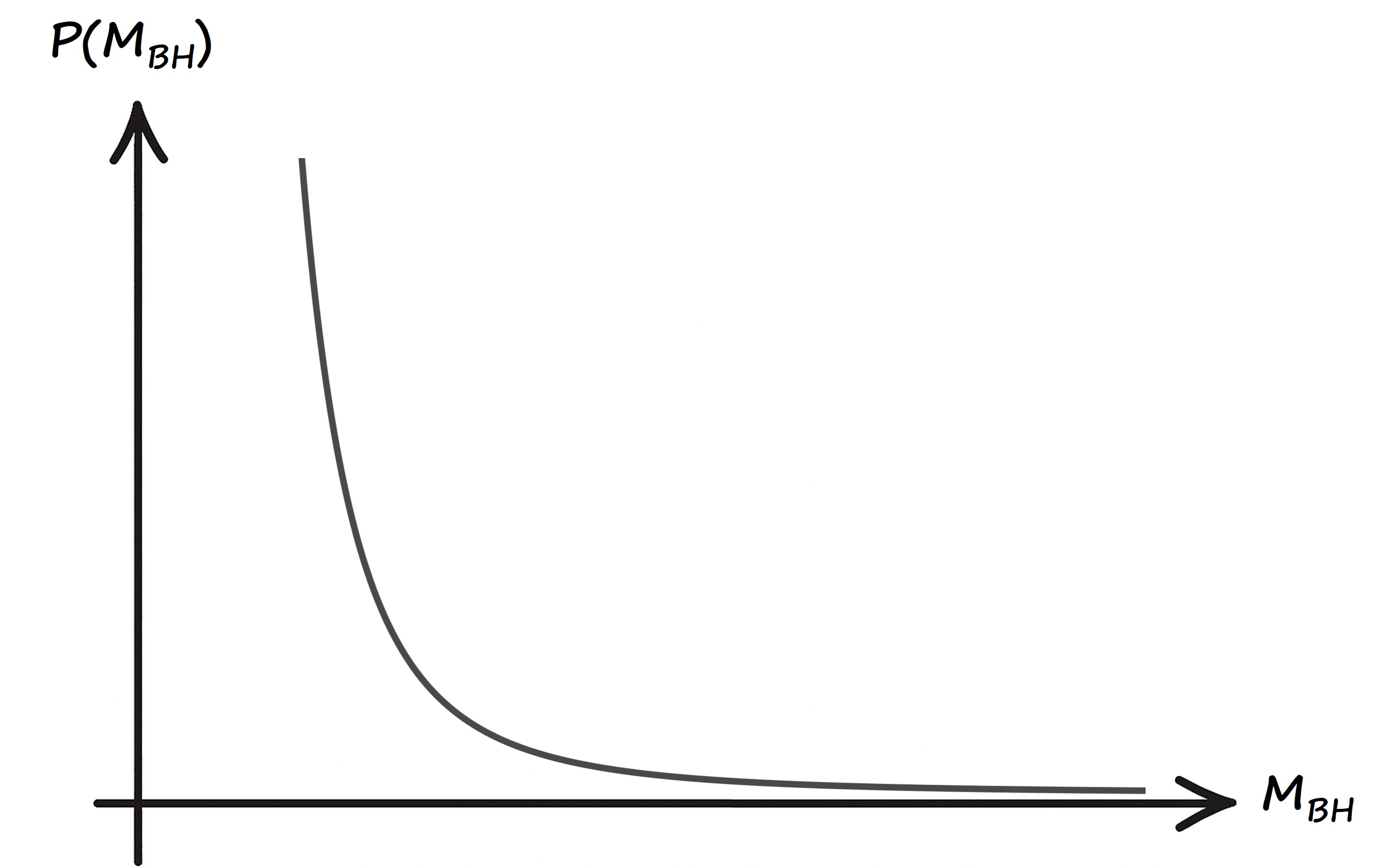}
  \caption{Plot of radiated power versus black hole mass.}
\end{figure}

On the other hand, according to Einstein's mass--energy relation, the energy stored in the mass of the black hole is $E = M_{BH} c^2$. This expression allows us to write the radiated power as $P = E/\Delta t = M_{BH} c^2 / \Delta t$, where $\Delta t$ is the corresponding time interval:

\begin{equation}
\frac{M_{BH} c^2}{\Delta t} = \frac{c^6 \hbar}{15360\pi G^2 M_{BH}^2}.
\end{equation}

Solving for $\Delta t$, we obtain an estimate of the evaporation time of a black hole, defined as the time it takes for its mass to decrease to zero:

\begin{equation}
\Delta t = \frac{15360\pi G^2 M_{BH}^3}{\hbar c^4}.
\end{equation}

In this expression, $M_{BH}$ is the initial mass of the black hole. More detailed calculations show that the evaporation time is one third of the previous value (see Appendix):

\begin{equation}
t_{\text{ev}} = \frac{5120\pi G^2 M_{BH}^3}{\hbar c^4} \simeq 6.6 \times 10^{74}\,\text{s} \left(\frac{M_{BH}}{M_{\odot}}\right)^3,
\end{equation}

where $t_{\text{ev}}$ has been expressed in units of solar mass, and the equation is given in seconds (s). Figure~5 shows the graph of this equation, where it is evident that the evaporation time increases rapidly with increasing initial mass of the black hole. We can obtain a more general expression by considering the evaporation times for two arbitrary mass values. If $M_i$ is the initial mass of the black hole at an initial time $t_i$, and $M_f$ is the final mass at a time $t_f$, where $M_i > M_f$ and $t_i < t_f$, then from Eq.~(11) we obtain the following expressions:

\begin{equation}
t_i = \frac{5120\pi G^2 M_i^3}{\hbar c^4}, \quad t_f = \frac{5120\pi G^2 M_f^3}{\hbar c^4}.
\end{equation}

The interval $|-\Delta t| = |t_i - t_f| \equiv \tau$ corresponds to the time it takes for a black hole to reduce its mass from $M_i$ to $M_f$ due to evaporation:

\begin{equation}
\tau = \frac{5120\pi G^2}{\hbar c^4}(M_i^3 - M_f^3) = \frac{5120\pi G^2 M_i^3}{\hbar c^4} \left[1 - \left(\frac{M_f}{M_i}\right)^3\right] = t_{\text{ev}} \left[1 - \left(\frac{M_f}{M_i}\right)^3\right].
\end{equation}

We see that when the final mass is zero ($M_f = 0$), then $\tau = t_{\text{ev}}$. This equation shows that the time required for a black hole to reduce its mass, for example, to $10\%$ of its initial value, is almost equal to the evaporation time, since for $M_f = M_i/10$ in Eq.~(13) we have $\tau \simeq 0.999\,t_{\text{ev}}$. This implies that most of the evaporation time elapses during the early stages, while the final phase is extremely rapid and highly energetic. \\

\begin{figure}[H]
  \centering
    \includegraphics[width=0.3\textwidth]{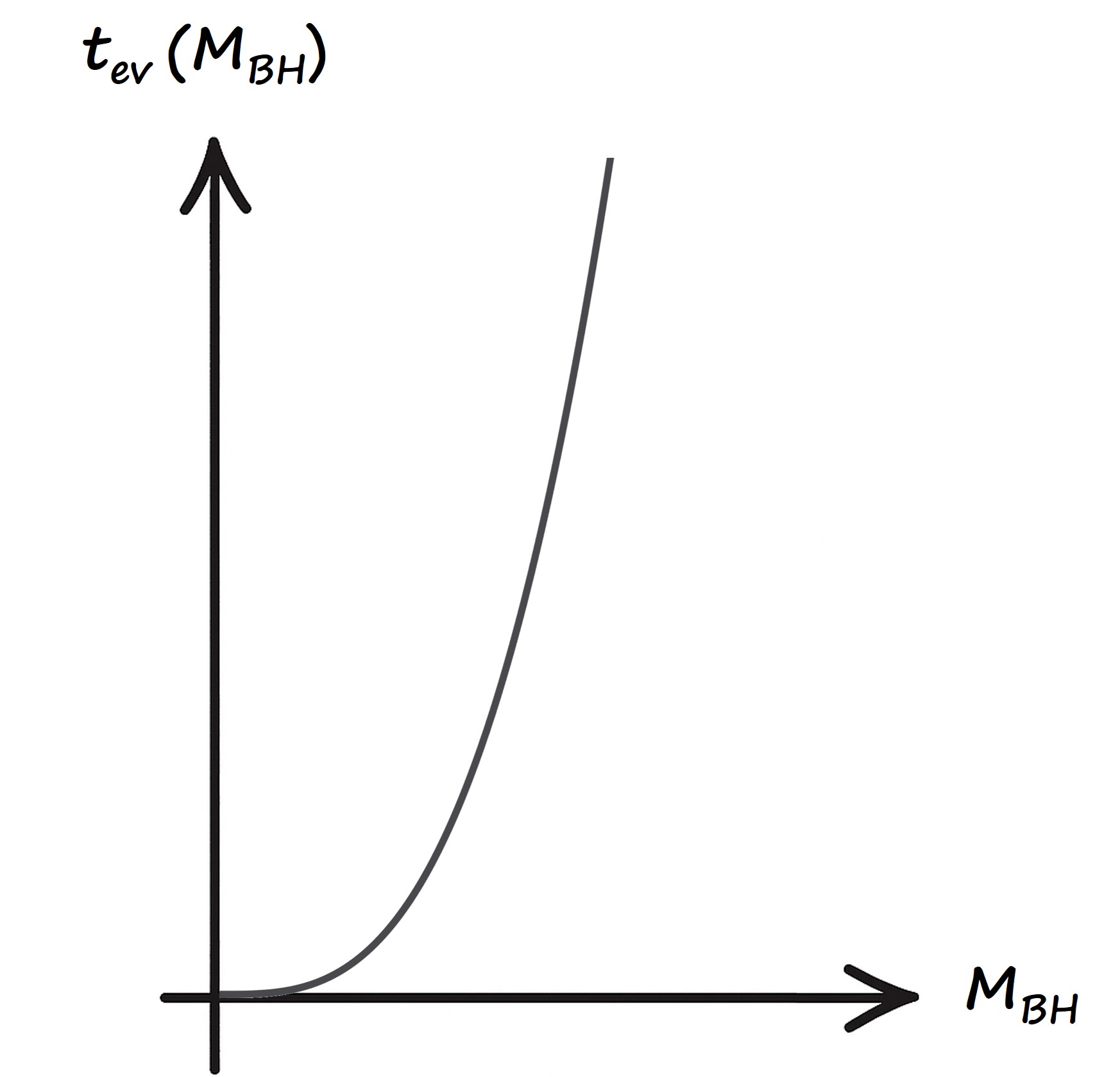}
  \caption{Plot of evaporation time versus black hole mass.}
\end{figure}

Let us now consider a specific example, assuming that the initial mass of a Schwarzschild black hole is equal to the solar mass. Taking $M_{BH} = M_{\odot}$ in Eq.~(11), we find that $t_{\text{ev}} \simeq 6.6 \times 10^{74}\,\text{s}$. This quantity is unimaginably large, since it is $58$ orders of magnitude greater than the age of the universe, estimated to be $\sim 10^{17}\,\text{s}$. Therefore, there is no possibility of detecting the evaporation process in stellar-mass black holes, which, as previously noted, are the lightest for which there is astronomical evidence.\\

However, even if astronomers had an infinite amount of time to observe black hole evaporation, there is another difficulty that makes detection impossible. A black hole produces a net outward flux of radiation only if it emits more particles than it absorbs. Nevertheless, since the cosmic microwave background (CMB) radiation (the primordial radiation that fills the universe) has a temperature $T_{\text{CMB}} \simeq 2.73\,\text{K}$ [25], which is much higher than the typical Hawking temperature of a stellar black hole, $10^{-8}\,\text{K}$, there must be a net flux of radiation directed inward across the event horizon. This implies an increase in $M_{BH}$ and a decrease in $T_H$, and therefore stellar black holes are not currently evaporating. If, as observational evidence suggests, the universe is destined to expand forever, there will come a time when $T_{\text{CMB}} < T_H$, and as a result, the mass of stellar black holes will begin to decrease. However, since $T_{\text{CMB}}$ is inversely proportional to the scale factor $R$, which is a dimensionless measure of the expansion rate of the universe, for $T_{\text{CMB}} < T_H$ the universe would have to expand by at least a factor of $\sim 10^8$. This result is consistent with what was noted in the previous section, where we concluded that, in the case of a stellar black hole, $T_H$ is undetectable\footnote{The mathematical relation can be written as $T_{\text{CMB}1}/T_{\text{CMB}2} = R_2/R_1$. If $T_{\text{CMB}1} = 2.73\,\text{K}$, $T_{\text{CMB}2} \sim 10^{-8}\,\text{K}$, and $R_1 = 1$ is the scale factor for the current size of the universe, then $R_2 \sim 10^8$.}.\\

Therefore, it appears that astronomy has little to contribute to the verification of Hawking's theories. However, specialists have analyzed other scenarios that allow the study of Hawking's results through experiments carried out on Earth. Due to space limitations, we cannot analyze this topic here. The reader can find a non-technical discussion in [14].

\section{What types of particles are emitted during evaporation?}

The first physicist to perform detailed numerical calculations on the types of particles that constitute Hawking radiation was Don Page in 1976 [26]. Although Page's work is highly technical, the basic idea underlying his calculations is simple. From elementary thermodynamic considerations, we know that the characteristic energy of a particle in Hawking radiation is $k T_H$. Thus, in practice, a particle of energy $E$ or mass $m = E/c^2$ can only be emitted if

\begin{equation}
E \lesssim k T_H = \frac{\hbar c^3}{8\pi G M_{BH}} \sim 10^{-12}\,\text{eV} \left(\frac{M_{\odot}}{M_{BH}}\right),
\end{equation}

where $1\,\text{eV} = 1.60 \times 10^{-19}\,\text{J}$. This equation shows that as the black hole evaporates and $M_{BH}$ decreases, the mass--energy of the emitted particles becomes progressively larger.\\

\begin{figure}[H]
  \centering
    \includegraphics[width=0.4\textwidth]{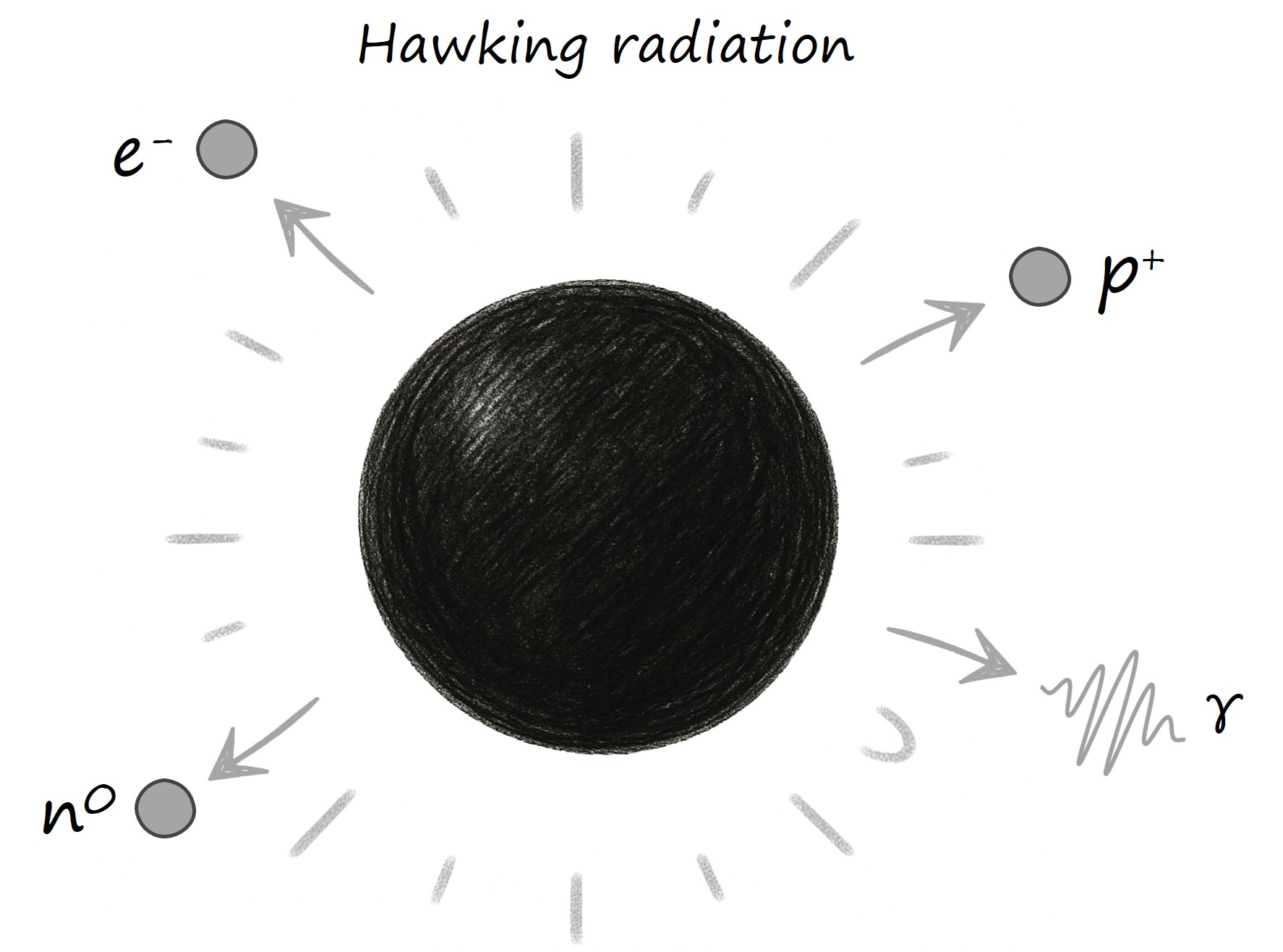}
  \caption{As a black hole evaporates, the mass--energy of the particles composing Hawking radiation can span the entire range of existing particles.}
\end{figure}

Let us consider a black hole of solar mass, $M_{BH} = M_{\odot}$ ($R_S = 3\,\text{km}$). In this case, the maximum energy of the emitted particles is $E \sim 10^{-12}\,\text{eV}$, which corresponds to radio-wave photons. If, after a long evaporation process, the black hole has reduced its mass to, for example, $M_{BH} \sim 10^{10}\,\text{kg}$ ($R_S \sim 10^{-17}\,\text{m}$), the maximum mass--energy of the emitted particles will be $E/c^2 \sim 10^{8}\,\text{eV}/c^2$, which is of the order of the mass of a proton, a neutron, or a hydrogen atom.\\

This indicates that, as the black hole evaporates, the mass–energy of the emitted particles can span the entire range of known particles (Fig.~6). However, this emission is not determined solely by the value of $M_{BH}$, since, due to conservation laws, certain quantities must be preserved during the process, such as energy, electric charge, and the lepton and baryon numbers. The way in which this occurs has been described in non-technical terms by Hawking himself in his popular science books [3,22].

\section{The final stage of the evaporation process}

Although so far we have assumed that, after evaporation, the black hole disappears completely, there is a serious problem with this assumption that must be carefully analyzed. Since, according to Eqs.~(3) and (4), $T_H \sim 1/M_{BH}$, the Hawking temperature should become infinite in the final moments of the black hole's life, when $M_{BH}$ tends to zero. This suggests that Eq. (4) does not have general validity, indicating that a more complete theory is required to describe the final stages of black hole evaporation. Some specialists consider that the solution to this problem requires assuming that there exists a minimum mass below which the black hole cannot continue evaporating. This is the Planck mass [21]:

\begin{equation}
m_P \equiv \left(\frac{\hbar c}{G}\right)^{1/2} \simeq 2.2 \times 10^{-8}\,\text{kg}.
\end{equation}

As the reader can easily verify, if in Eq.~(1) we take $M_{BH} = m_P$, apart from a factor of $2$ one obtains $R_S \simeq l_P$, where $l_P$ is the Planck length [21]:

\begin{equation}
l_P \equiv \left(\frac{G\hbar}{c^3}\right)^{1/2} \simeq 1.6 \times 10^{-35}\,\text{m}.
\end{equation}

This quantity is considered the smallest unit of distance to which physical meaning can be assigned [9]. To appreciate how small this value is, recall that the typical radius of a proton is $\sim 10^{-15}\,\text{m}$, which is a factor of $10^{20}$ larger than $l_P$. These ideas suggest that the final result of evaporation is a black hole remnant with mass $m_P$ and radius $l_P$.\\

At present, there is no consistent physical theory that supports the existence of a black hole remnant with mass $m_P$ and radius $l_P$, nor is there empirical evidence in favor of this idea. Therefore, it is not known exactly what happens to a black hole when its mass approaches $m_P$. However, for reasons of consistency with other physical theories that have been extensively confirmed, specialists agree that Eqs.~(4) and (11) are correct, provided that $M_{BH} \geq m_P$ and $R_S \geq l_P$. The resolution of this problem is one of the great challenges of contemporary physics.

\section{Final remarks}

We have seen that it is possible to acquire a qualitative understanding of the evaporation process by analyzing simple physical scenarios and performing heuristic calculations. Therefore, it is expected that this article provides the reader with a first approach to the study of black hole evaporation. Undoubtedly, many questions remain to be answered, as this is a topic at the frontier of physics and is actively investigated. The main educational contribution of this work is to bring an active research problem into the classroom. By integrating concepts from different areas of physics without resorting to advanced mathematics, the article promotes an early approach of students to cutting-edge topics.

\section*{Appendix: More detailed derivation of the evaporation time}

Let us begin with the Stefan--Boltzmann law for the power radiated by a blackbody, which, in the case of a black hole of mass $M$, can be written as

\begin{equation*}
\frac{dE}{dt} = \sigma A T_H^4,
\end{equation*}

where $A$ is the area of the event horizon, which is given by $A = 16\pi G^2 M^2/c^4$. Substituting this value of $A$ into the first equation and recalling that $\sigma = \pi^2 k^4 / (60 \hbar^3 c^2)$, we obtain

\begin{equation*}
\frac{dE}{dt} = \frac{c^6 \hbar}{15360\pi G^2 M^2}.
\end{equation*}

On the other hand, according to relativistic physics, the total energy of the black hole is $E = M c^2$, and therefore the radiated power can also be written as

\begin{equation*}
\frac{dE}{dt} = -c^2 \frac{dM}{dt}.
\end{equation*}

The minus sign indicates that a decrease in mass implies emission of energy. Eliminating the derivative $dE/dt$ between the last two equations, we obtain

\begin{equation*}
-\frac{dM}{dt} = \frac{c^4}{15360\pi G^2 \hbar M^2}.
\end{equation*}

This differential equation can be solved by direct integration. If $M_i$ is the initial mass of the black hole at time $t_i$, and $M_f$ is the final mass at time $t_f$, we obtain:

\begin{equation*}
- \int_{M_i}^{M_f} M^2 \, dM = \int_{t_i}^{t_f} \frac{c^4}{15360\pi G^2 \hbar} \, dt.
\end{equation*}

Integrating both sides of the equality, we recover Eq.~(13):

\begin{equation*}
\tau \equiv t_f - t_i = \frac{5120\pi G^2}{\hbar c^4}(M_i^3 - M_f^3) = t_{\text{ev}} \left[1 - \left(\frac{M_f}{M_i}\right)^3\right].
\end{equation*}

This is the time required for a black hole to reduce its mass from $M_i$ to $M_f$. By definition, the evaporation time occurs when $M_f = 0$, which allows us to recover Eq.~(11):

\begin{equation*}
t_{\text{ev}} = \frac{5120\pi G^2}{\hbar c^4} M_i^3.
\end{equation*}

\section*{References}

\begin{enumerate}
\item K. Ferguson, \textit{Stephen Hawking: An Unfettered Mind}, St. Martin’s Press, New York, 2017.
\item S.W. Hawking, ``Black Holes and Thermodynamics,'' \textit{Physical Review D} \textbf{13} (1976) 191--197.
\item S.W. Hawking, \textit{A Brief History of Time}, Bantam Books, New York, 1998.
\item S.V. Kontomaris, A. Malamou, ``A Presentation of the Black Hole Stretching Effect,'' \textit{Physics Education} \textbf{53} (2018) 015010.
\item P.J. Riggs, ``Black Holes: Basic Concepts and Popular Misconceptions,'' \textit{Physics Education} \textbf{54} (2019) 065015.
\item M. Tuveri, A. Steri, D. Fadda, ``Using Storytelling to Foster the Teaching and Learning of Gravitational Waves Physics at High-School,'' \textit{Physics Education} \textbf{59} (2024) 045031.
\item D. Lincoln, A.L. Stuver, ``Ripples in Reality,'' \textit{The Physics Teacher} \textbf{54} (2016) 398--403.
\item J. Pinochet, ``Hawking Temperature: An Elementary Approach Based on Newtonian Mechanics and Quantum Theory,'' \textit{Physics Education} \textbf{51} (2016) 015010.
\item J. Pinochet, ``The Hawking Temperature, the Uncertainty Principle and Quantum Black Holes,'' \textit{Physics Education} \textbf{53} (2018) 065004.
\item J. Pinochet, ``Black holes ain't so black: An introduction to the great discoveries of Stephen Hawking,'' \textit{Physics Education} \textbf{54} (2019) 035014.
\item J. Pinochet, ``Five misconceptions about black holes,'' \textit{Physics Education} \textbf{54} (2019) 055003.
\item J. Pinochet, ``Hawking for beginners: A dimensional analysis activity to perform in the classroom,'' \textit{Physics Education} \textbf{55} (2020) 045018.
\item J. Pinochet, ``Three easy ways to the Hawking temperature,'' \textit{Physics Education} \textbf{56} (2021) 053001.
\item J. Pinochet, ``Hawking for everyone: Commemorating half a century of an unfinished scientific revolution,'' \textit{Physics Education} \textbf{59} (2024) 055001.
\item J. Pinochet, ``Hawking in a thousand words,'' \textit{Physics Education} \textbf{60} (2025) 063003.
\item M.C. LoPresto, ``Some Simple Black Hole Thermodynamics,'' \textit{The Physics Teacher} \textbf{41} (2003) 299--301.
\item A. Einstein, ``Die Grundlage der allgemeinen Relativitätstheorie,'' \textit{Annalen der Physik} \textbf{354} (1916) 769--822.
\item H. Gutfreund, J. Renn, \textit{The Road to Relativity: The History and Meaning of Einstein’s ``The Foundation of General Relativity''}, Princeton University Press, 2015.
\item B. Schutz, \textit{Gravity from the Ground Up}, Cambridge University Press, Cambridge, 2003.
\item J.P. Luminet, \textit{Black Holes}, Cambridge University Press, Cambridge, 1999.
\item P.A. Tipler, R.A. Llewellyn, \textit{Modern Physics}, W. H. Freeman and Company, New York, 2012.
\item S.W. Hawking, \textit{The Universe in a Nutshell}, Bantam Books, New York, 2001.
\item R.A. Serway, J.W. Jewett, \textit{Physics for Scientists and Engineers with Modern Physics}, Thomson, Cengage Learning, 2010.
\item R. Resnick, D. Halliday, K.S. Krane, \textit{Physics}, 4th ed., John Wiley and Sons, New York, 1992.
\item D. Maoz, \textit{Astrophysics in a Nutshell}, Princeton University Press, Princeton, 2016.
\item D.N. Page, ``Particle emission rates from a black hole: Massless particles from an uncharged, non-rotating hole,'' \textit{Physical Review D} \textbf{13} (1976) 198--206.
\end{enumerate}

\end{document}